\documentclass[fleqn,12pt,twoside]{article}
\usepackage{espcrc1}

\usepackage{graphicx}
\usepackage[figuresright]{rotating}

\newcommand{\AmS}{{\protect\the\textfont2
  A\kern-.1667em\lower.5ex\hbox{M}\kern-.125emS}}
\newcommand{\be}{\begin{equation}}
\newcommand{\ee}{\end{equation}}
\newcommand{\bea}{\begin{eqnarray}}
\newcommand{\eea}{\end{eqnarray}}

\title{Effective Theory of the Color-Flavor-Locked Phase }

\author{Thomas Sch\"afer\address{Department of Physics and Astronomy, 
     State University of New York, 
     Stony Brook, NY 11794-3800 
     and Riken-BNL Research Center, Brookhaven National 
     Laboratory, Upton, NY 11973}}
       
\begin{document}

\maketitle

\begin{abstract}
We explain how an effective theory of the CFL phase can be
used to study the effect of the strange quark mass on the 
ground state and the excitation spectrum. We also apply 
the effective theory to the problem of neutrino emission 
from a CFL quark core inside a neutron star. 
\end{abstract}

\section{Introduction}
\label{sec_intro}

  What is the ground state of ordinary hadronic matter if it
is compressed to baryon densities several times larger than 
nuclear matter saturation density? This theoretical question,
which has important applications to the physics of neutron 
stars, has led to the discovery of several novel and unusual 
phases of QCD. At zero baryon density chiral symmetry 
is broken by a condensate of quark-anti-quark pairs. If the 
baryon density is large then condensation in the quark-anti-quark 
channel is suppressed. Instead, attractive interactions between
two quarks in a color anti-symmetric wave function lead to 
diquark condensation and color superconductivity 
\cite{Bailin:1984bm,Alford:1998zt,Rapp:1998zu}. 

  The global symmetries of color superconducting quark matter
depend on the density, the number of quark flavors, and their 
masses. In the physically relevant case of three light quark 
flavors a particularly symmetric phase exists, color-flavor-locked 
(CFL) quark matter \cite{Alford:1999mk}. This phase is believed to 
be the true ground state of strange quark matter at very large 
density. The CFL phase is characterized by the order parameter
\be
\label{cfl}
\langle q_{L,i}^a C q_{L,j}^b\rangle
 = -\langle q_{R,i}^a C q_{R,j}^b\rangle
 = \phi \left(\delta_i^a\delta_j^b-\delta_i^b\delta_j^a\right).
\ee
At realistic baryon densities flavor symmetry breaking 
due to the quark masses $m_s\neq m_d\neq m_u$ and 
non-zero lepton chemical potentials will lead to distortions
of the ideal CFL state
\cite{Alford:1999pa,Schafer:1999pb,Schafer:2000ew,Rajagopal:2001ff,Bedaque:2001je,Kaplan:2001qk,Buballa:2001gj,Kryjevski:2002ju,Alford:2002kj,Steiner:2002gx}.
In this contribution we show how to address this problem using the 
effective chiral theory of the CFL phase \cite{Casalbuoni:1999wu}.

\section{Chiral Effective Theory}
\label{sec_CFLchi}

 For excitation energies smaller than the gap the only 
relevant degrees of freedom are the Goldstone modes 
associated with the breaking of chiral symmetry and
baryon number. The interaction of the Goldstone modes
is described by the effective Lagrangian \cite{Casalbuoni:1999wu}
\bea
\label{l_cheft}
{\cal L}_{eff} &=& \frac{f_\pi^2}{4} {\rm Tr}\left[
 \nabla_0\Sigma\nabla_0\Sigma^\dagger - v_\pi^2
 \partial_i\Sigma\partial_i\Sigma^\dagger \right] 
 +\Big[ C {\rm Tr}(M\Sigma^\dagger) + h.c. \Big] \\ 
 & & \hspace*{-1cm}\mbox{} 
     +\Big[ A_1{\rm Tr}(M\Sigma^\dagger)
                        {\rm Tr} (M\Sigma^\dagger) 
     + A_2{\rm Tr}(M\Sigma^\dagger M\Sigma^\dagger)   
     + A_3{\rm Tr}(M\Sigma^\dagger){\rm Tr} (M^\dagger\Sigma)
         + h.c. \Big]+\ldots . 
 \nonumber 
\eea
Here $\Sigma=\exp(i\phi^a\lambda^a/f_\pi)$ is the chiral field,
$f_\pi$ is the pion decay constant and $M$ is the mass matrix. 
The field $\phi^a$ describes pion, kaon, and eta collective 
modes in the CFL phase. The microscopic nature of these 
modes is not relevant for our discussion. It is nevertheless
useful to have a physical picture of these excitations, see
Fig.~1a. In the CFL phase the flavor and color orientation 
of the left handed condensate $X_i^a = \epsilon_{ijk}
\epsilon^{abc}\langle (\psi_L)_j^b C(\psi_L)_k^c \rangle$ 
are locked, $X_i^a\sim \delta_i^a$. The same is true for
the right handed condensate $Y_i^a = \epsilon_{ijk}\epsilon^{abc}
\langle (\psi_R)_j^b C(\psi_R)_k^c \rangle$. Low energy 
excitations of the CFL phase correspond to small fluctuations
of $X$ and $Y$ around their equilibrium values. Because color 
gauge invariance is broken colored excitations acquire a
mass via the Higgs mechanism. The true low energy modes
are color neutral fluctuations of $X$ relative to $Y$,
parameterized by $\Sigma=XY^\dagger$. For example, a
low energy mode with the quantum numbers of the $K^0$ 
is given by $K^0\sim \epsilon^{abc}\epsilon^{ade} 
(\bar{u}^b_RC\bar{s}_R^c) (d^d_LCu^e_L)$. 

\begin{figure}[t]
\includegraphics[width=11cm]{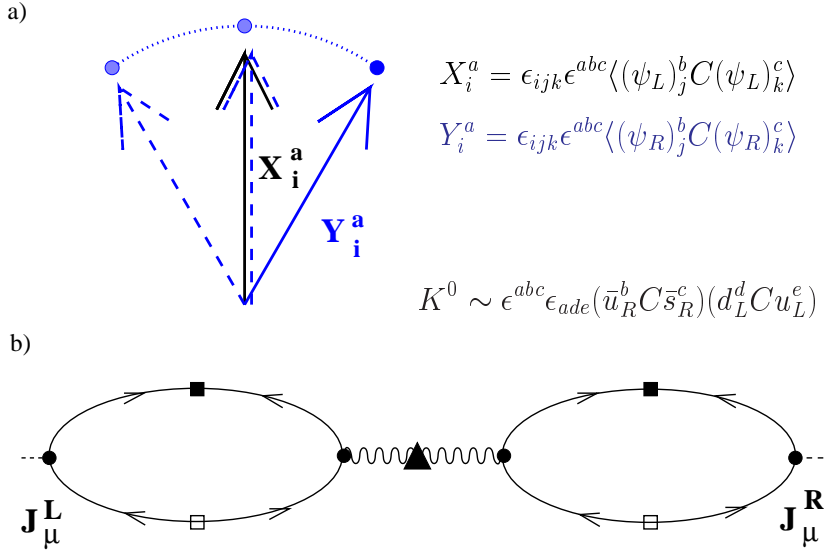}
\caption{\label{fig_gb}
Figure a) shows schematically how to construct 
fluctuations of the CFL order parameter that have the
quantum numbers of the $K^0$. Fig. b) shows the leading 
order contribution to the $\langle j_\mu^L j_\mu^R\rangle$
correlation function which determines $f_\pi^2$ in the CFL
phase. The open and solid squares are insertions of $
\langle \psi_{L,R}\psi_{L,R}\rangle$ and $\langle 
\bar{\psi}_{L,R}\bar{\psi}_{L,R}\rangle$. Note that the 
graph involves $XY^\dagger X^\dagger Y \sim \Sigma 
\Sigma^\dagger$.}
\end{figure}

 At very high baryon density the effective coupling is weak
and the coefficients $f_\pi^2,C,A_i$ can be determined in 
perturbative QCD. The pion decay constant is given by (see
Fig.~1b) \cite{Son:1999cm} 
\be
\label{f_pi}
f_\pi^2 = \frac{21-8\log(2)}{18} 
  \left(\frac{p_F^2}{2\pi^2} \right) .
\ee 
The coefficient $C$ is related to instantons and was 
computed in \cite{Schafer:2002ty}. At large baryon density
$C\sim (\Lambda_{QCD}/p_F)^8$ and the linear mass term 
is not important. The coefficients $A_i$ of the quadratic
mass terms are given by \cite{Son:1999cm,Schafer:2001za} 
\be
\label{A_i}
 A_1= -A_2 = \frac{3\Delta^2}{4\pi^2}, 
\hspace{1cm} A_3 = 0.
\ee
Finally, the covariant derivative $\nabla_\mu\Sigma$ was
determined in \cite{Bedaque:2001je}. The temporal
component $\nabla_0\Sigma$ contains the quark mass
matrix, 
\be
\label{mueff}
 \nabla_0\Sigma = \partial_0 \Sigma 
 + i \left(\frac{M M^\dagger}{2p_F}\right)\Sigma
 - i \Sigma\left(\frac{ M^\dagger M}{2p_F}\right) .
\ee
We note that equ.~(\ref{mueff}) is completely fixed
by the symmetries of the theory. In addition to that,
using the power counting proposed in \cite{Bedaque:2001je}
we find that the low energy constants $A_i$ are of 
natural magnitude. This suggests that even though 
equs.~(\ref{f_pi}-\ref{mueff}) were obtained using 
weak coupling methods, the results are more general.

\section{Kaon Condensation}
\label{sec_kcond}

 The effective chiral lagrangian equ.~(\ref{l_cheft}) 
determines the masses and interactions of the CFL
Goldstone bosons. In addition to that, the effective theory 
also determines the structure of the ground state as a function 
of the quark masses and external fields that couple to the 
Goldstone modes \cite{Bedaque:2001je,Kaplan:2001qk}. The 
effective potential is given by
\be
\label{veff}
V_{eff}(\Sigma) = \frac{f_\pi^2}{2} {\rm Tr}\left[
 X_L\Sigma X_R \Sigma^\dagger \right] 
   + A\Big[ {\rm Tr}(M\Sigma^\dagger)
                        {\rm Tr} (M\Sigma^\dagger) 
       -{\rm Tr}(M\Sigma^\dagger M\Sigma^\dagger)   
         + h.c. \Big] , 
 \nonumber 
\ee
with $X_L=MM^\dagger/(2p_F)$ and $X_R=M^\dagger M/(2p_F)$.
For small $M$ the minimum of $V_{eff}$ is at $\Sigma=1$. 
If the mass is increased a number of different phases
can occur. In practice we are mostly interested in the 
case $m_s\gg m_d\simeq m_u$. In this case there is an
instability towards kaon condensation. The simplest ansatz 
for a $K^0$ condensed ground state is given by $\Sigma = 
\exp\left(i\alpha \lambda_4\right)$. With this ansatz 
the vacuum energy is 
\be 
\label{k0+_V}
 V(\alpha) = -f_\pi^2 \left( \frac{1}{2}\left(\frac{m_s^2-m^2}{2p_F}
   \right)^2\sin(\alpha)^2 + (m_{K}^0)^2(\cos(\alpha)-1)
   \right),
\ee
where $(m_K^0)^2= (4A/f_\pi^2)m_{u,d} (m_{u,d}+m_s)$ is 
the $O(M^2)$ kaon mass in the limit of exact isospin 
symmetry. Minimizing the vacuum energy we obtain $\alpha=0$ 
if $m_s^2/(2p_F)<m_K^0$ and $\cos(\alpha)=(m_K^0)^2/\mu_s^2$ 
with $\mu_s=m_s^2/(2p_F)$ if $\mu_s>m_K^0$. The hypercharge 
density is given by
\be 
n_Y = f_\pi^2 \mu_s \left( 1 -\frac{(m_K^0)^4}{\mu_s^4}\right).
\ee
We observe that within the range of validity of the effective 
theory, $\mu_s<\Delta$, the hypercharge density satisfies 
$n_Y<\Delta p_F^2/(2\pi^2)$. This means that the number of 
condensed kaons is bounded by the number of particles contained 
within a strip of width $\Delta$ around the Fermi surface. It 
also implies that near the unlocking transition, $\mu_s\sim
\Delta$, the CFL state is significantly modified. In this
regime, of course, we can no longer rely on the effective 
theory and a more microscopic calculation is necessary. 

\section{Summary and Outlook}
\label{sec_sum}

 We have studied the ground state of CFL quark matter for 
non-zero quark masses. We have argued that there is a new 
scale $m_s^2/(2p_F)\ll (\Delta/p_F)$ which corresponds to 
the onset of kaon condensation. For larger values of the 
strange quark mass, $m_s^2/(2p_F)\sim 1$, color-flavor-locking
breaks down. Using weak coupling methods, we can determine the 
critical $m_s$ for kaon condensation to occur. We find $m_s
|_{crit}\simeq 3 (m_{u,d}\Delta^2)^{1/3}$. This result suggests 
that for values of the strange quark mass and the gap that are 
relevant to compact stars CFL matter is likely to support 
a kaon condensate. 

 The effective lagrangian equ.~(\ref{l_cheft}) can also be 
used to compute many physical properties of the CFL phase. 
In recent work \cite{Reddy:2002xc,Jaikumar:2002vg} we considered 
the specific heat, neutrino emission and neutrino scattering in 
the CFL phase. At energies $E<\Delta$ these processes are 
dominated by Goldstone modes. We find that the long term 
emissivity of the CFL phase is very low, and that the CFL 
phase contributes little to the long term cooling of a 
neutron star with a CFL core. Interaction in the CFL
phase are important in determining the neutrino 
opacity at early times \cite{Reddy:2002xc}.


\begin{thebibliography}{8.}
\addcontentsline{toc}{section}{References}

\bibitem{Bailin:1984bm}
D.~Bailin and A.~Love,
Phys.\ Rept.\  {\bf 107}, 325 (1984).
 
\bibitem{Alford:1998zt}
M.~Alford, K.~Rajagopal and F.~Wilczek,
Phys.\ Lett.\  {\bf B422}, 247 (1998).
 
\bibitem{Rapp:1998zu}
R.~Rapp, T.~Sch{\"a}fer, E.~V.~Shuryak and M.~Velkovsky,
Phys.\ Rev.\ Lett.\  {\bf 81}, 53 (1998).

\bibitem{Alford:1999mk}
M.~Alford, K.~Rajagopal and F.~Wilczek,
Nucl.\ Phys.\  {\bf B537}, 443 (1999).

\bibitem{Alford:1999pa}
M.~Alford, J.~Berges and K.~Rajagopal,
Nucl.\ Phys.\  {\bf B558}, 219 (1999).
 
\bibitem{Schafer:1999pb}
T.~Sch{\"a}fer and F.~Wilczek,
Phys.\ Rev.\  {\bf D60}, 074014 (1999).

\bibitem{Schafer:2000ew}
T.~Sch{\"a}fer,
Phys.\ Rev.\ Lett.\ {\bf 85}, 5531 (2000).

\bibitem{Rajagopal:2001ff}
K.~Rajagopal and F.~Wilczek,
Phys.\ Rev.\ Lett.\  {\bf 86}, 3492 (2001).

\bibitem{Bedaque:2001je}
P.~F.~Bedaque and T.~Sch{\"a}fer,
Nucl.\ Phys.\ A {\bf 697}, 802 (2002)
[hep-ph/0105150].

\bibitem{Kaplan:2001qk}
D.~B.~Kaplan and S.~Reddy,
Phys.\ Rev.\ D {\bf 65}, 054042 (2002)
[hep-ph/0107265].

\bibitem{Buballa:2001gj}
M.~Buballa and M.~Oertel,
Nucl.\ Phys.\ A {\bf 703}, 770 (2002)
[hep-ph/0109095].

\bibitem{Kryjevski:2002ju}
A.~Kryjevski and T.~Norsen,
Phys.\ Rev.\ D {\bf 66}, 034010 (2002)
[hep-ph/0202208].

\bibitem{Alford:2002kj}
M.~Alford and K.~Rajagopal,
JHEP {\bf 0206}, 031 (2002)
[hep-ph/0204001].

\bibitem{Steiner:2002gx}
A.~W.~Steiner, S.~Reddy and M.~Prakash,
preprint, hep-ph/0205201.

\bibitem{Casalbuoni:1999wu}
R.~Casalbuoni and D.~Gatto,
Phys.\ Lett.\ {\bf B464}, 111 (1999).

\bibitem{Son:1999cm}
D.~T.~Son and M.~Stephanov, 
Phys.\ Rev.\ {\bf D61}, 074012 (2000), 
[hep-ph/9910491];
erratum: hep-ph/0004095.

\bibitem{Schafer:2002ty}
T.~Sch\"afer,
Phys.\ Rev.\ D {\bf 65}, 094033 (2002)
[hep-ph/0201189].

\bibitem{Schafer:2001za}
T.~Sch\"afer,
Phys.\ Rev.\ D {\bf 65}, 074006 (2002)
[hep-ph/0109052].

\bibitem{Reddy:2002xc}
S.~Reddy, M.~Sadzikowski and M.~Tachibana,
preprint, nucl-th/0203011.

\bibitem{Jaikumar:2002vg}
P.~Jaikumar, M.~Prakash and T.~Sch\"afer,
preprint, astro-ph/0203088.


\end{thebibliography}
\end{document}